\documentclass[prl, twocolumn]{revtex4}
\usepackage{graphicx}
\usepackage{dcolumn}
\usepackage{bm}

\def\be{\begin{equation}}
\def\ee{\end{equation}}
\def\bea{\begin{eqnarray}}
\def\eea{\end{eqnarray}}
\newcommand{\eref}[1]{(\ref{#1})}

\newcommand\refeq[1]{Eq.~(\ref{eq:#1})}

\begin{document}

\author{Lida Ansari}
\author{Baruch Feldman}
\email{baruchf@alum.mit.edu}
\author{Giorgos Fagas}
\author{Jean-Pierre Colinge}
\author{James C.~Greer}
\affiliation{Tyndall National Institute, University College Cork, Cork, Ireland}

\title{Simulations of gated Si nanowires and 3-nm junctionless transistors }

\date{\today}

\begin{abstract}
Inspired by recent experimental realizations and theoretical simulations of thin silicon nanowire-based devices, we perform predictive 
first-principles simulations of junctionless gated Si nanowire transistors. 
Our primary predictions are that Si-based transistors are physically possible without major changes in design philosophy at scales of $\sim$1 nm wire diameter and $\sim$3 nm gate length, and that the junctionless transistor \cite{Junctionless, Nima} may be the only physically sensible design at these length scales.  We also present investigations into atomic-level design factors such as 
dopant positioning and concentration. 
\end{abstract}

\maketitle


As the semiconductor technology roadmap nears its end, more and more fundamental changes are becoming necessary to design transistor devices. Short-channel effects \cite{ChanLength, JP-book, Singapore} degrade subthreshold slope, aggravate drain-induced barrier lowering (DIBL), and limit overall performance.  In response, designs using more gates and thinner channels to enhance gating control and alleviate these effects are becoming popular \cite{ChanLength, JP-book, Singapore}.  

Recently, {\em junctionless} nanowire transistors were fabricated with a trigate electrode structure \cite{Junctionless}.  These nanowire transistors have a thickness of a few nm and channel length of 1 $\rm{\mu}$m.  This design, essentially a ``gated resistor'' that turns off by pinch-off when gate voltage is applied, avoids the difficulty of fabricating ultrashallow junctions (as in classic MOSFETs) at nanometer length scales \cite{Junctionless, Nima}.  Moreover, previous semi-classical simulations indicate it has better short-channel characteristics than comparable trigate MOSFETs \cite{Nima}.

In this letter, we continue our previous efforts \cite{Georgios, Felipe} to understand transport in Si nanowires by simulating an atomic-scale device with a gating field and calculating its $I$-$V_{ds}$ characteristics.  
We present first-principles calculations of the response of doped 
junctionless silicon nanowire (SiNW) transistors to source-drain bias, $V_{ds}$, and gate voltage, $V_g$.  
We note our simulations are {\em predictive}, applying to devices both thinner and shorter than those currently achievable in the lab \cite{Junctionless} or by effective-mass calculation \cite{Nima}.  At such small scales, classic two-junction transistor designs are difficult to fabricate, and (because of dopant de-localization, as we will discuss) may not be physically possible.  Most 
importantly, we find the junctionless transistor device concept works at scales as small as wire diameter of $\sim$1 nm and gate length of $\sim$3 nm.  

\begin{figure}
\includegraphics[scale=0.15]{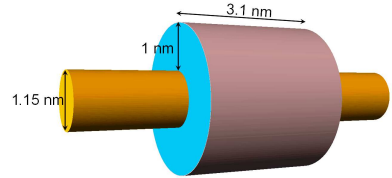}
\caption{\label{fig:device} Geometry of junctionless gate-all-around (GAA) SiNW devices simulated.  }
\end{figure}

\begin{figure}
\includegraphics[scale=0.35]{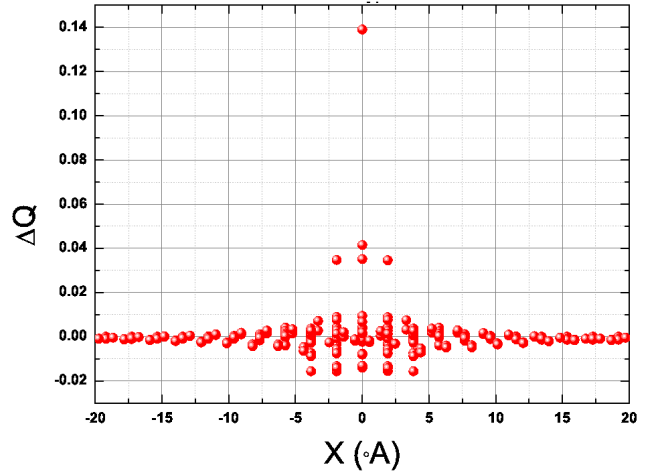}
\caption{\label{fig:Mulliken} Mulliken charge differences \eref{eq:Mul-diff} for n-doped {\em vs.} intrinsic Si nanowires as a function of position $x$ along the wire axis.  Shown here is a donor at the origin. }

\end{figure}

A typical structure of our simulated junctionless SiNW transistors is shown in Fig.~\ref{fig:device}.  As the name implies, these devices are uniformly doped throughout the wire from a macroscopic perspective \footnote{The entire channel region is comparable in length to the delocalized radius of the dopant electron or hole. See the discussion surrounding \refeq{radius-dopant} and Fig.~\ref{fig:Mulliken}. 
}.  As shown, the SiNWs have a gate-all-around (GAA) architecture.  In the actual devices realized experimentally \cite{Junctionless}, field effects from the work function of the gate cause the device to turn off at $V_g=0$ V.  But in principle, a junctionless device is a ``gated resistor'' that is {\em on} at $V_g=0$ V, as is the case in our simulations.  

\begin{figure}
\includegraphics[scale=0.4]{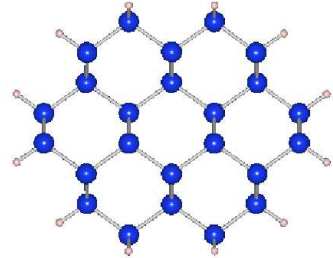}
\caption{\label{fig:NW-struct} Cross section of Si nanowire structures simulated.  }
\end{figure}

We used the [110]-oriented hydrogenated Si nanowire structures from previous work \cite{Georgios} (Fig.~\ref{fig:NW-struct}).  The wire diameter is $2 R_{NW} = 1.15$ nm.  

To find the electronic structure and Hamiltonian, we ran the self-consistent density functional tight-binding \cite{DFTB} code, DFTB$^+$.  DFTB$^+$ performs self-consistent electronic structure calculations in a tight-binding framework using parameters calculated from first-principles density functional theory (DFT) \cite{DFTB}.  This enabled us to simulate 
$\sim$800 atoms in our supercells.  

We simulated the gating field from a GAA structure by using point charges (the positions and charges of these are held fixed within the electronic-structure calculation).  We assembled the point charges in rings of radius $R_g = 1.6$ nm around the nanowire structure, typically containing 100 point charges per ring and spaced about 1 $\rm{\AA}$ apart along the wire axis.  
We used a gate length $L_g = 3.1$ nm.  The values of the point charges were fixed by the desired gate voltage, $V_g$. 
We modeled the oxide surrounding the nanowires by a continuum with hafnium oxide dielectric constant, $\epsilon_{\rm{HfO_2}} = 25$. 

To model doped nanowires, we inserted substitutional dopant atoms into the SiNW lattice.  We used Ga for a p-type dopant, and As for n-type.  Because of the relatively small supercells amenable to first-principles calculations, we used very high doping concentrations $N$ in the leads, typically $N = 8 \cdot 10^{20}$ cm$^{-3}$, about 10 times higher than in previous semi-classical simulations of junctionless transistors \cite{Nima}.  

Because our electronic structure calculations are based on DFT, all electrons are in their ground state, so in principle the dopants do not ionize.  However, even when setting the electronic temperature parameter in DFTB$^+$ to $T_e=0$ K, we found the lead Fermi levels and band structures output were consistent with many free carriers, 
\be
| E_F - E_d | \approx 350 \: \rm{meV}, 
\label{eq:dop-band}
\ee
with $E_d$ the edge of the dopant band, 
for such high $N$.  This behavior can be explained by modeling a dopant atom as a hydrogen-like system with effective electron mass $m^*$ and dielectric constant $\epsilon$ from Si \cite{AshcroftMermin}.  Then the typical localization radius of the dopant electron or hole is 
\be
R_{loc} = \frac{m}{m^*} \: \epsilon \: a_0,
\label{eq:radius-dopant}
\ee
with $a_0$ the Bohr radius.  
Using $m^*/m = 0.15$ for [110] SiNWs from our calculations \cite{Felipe} 
and the bulk value $\epsilon_{Si} = 11.7$, we find $R_{loc} = 4$ nm, even at 0 K.  
This is to compare to a dopant spacing of $\sim1$ nm along the wire.  

To understand this behavior better, we studied the Mulliken charge distributions for our doped and undoped SiNWs. 
Fig.~\ref{fig:Mulliken} 
shows the Mulliken charge differences, 
\be
\Delta Q^M_i \equiv Q^M_i - Q^{M,0}_i,
\label{eq:Mul-diff}
\ee
where $i$ is an atom index, $ Q^M_i$ is the Mulliken charge on atom $i$ in the n-doped wire, and  $Q^{M,0}_i$ is the Mulliken charge in the intrinsic wire. 
As shown in the figure, the donated electron de-localizes around the dopant atom with an exponential localization distance $R_{loc} = 1.5$ nm (for the wavefunction), in rough agreement with \refeq{radius-dopant}.  Further, the Mulliken charge differs from the intrinsic case throughout our supercell, confirming that at such high doping concentrations and in such a thin nanowire, dopants do not have to ionize to contribute to the channel's ``on'' conductivity.  
This behavior makes classical junctioned transistor designs with small gate lengths very difficult to achieve.  

Rurali {\em et al.}'s \cite{Rurali} calculations show that dopant levels are very deep in thin SiNWs, making dopants unlikely to ionize.  However, there is no contradiction with our finding in \refeq{dop-band} because our dopants are spaced so closely that the dopant band is highly curved.  Thus, we find $E_F \approx E_0$, where $E_0$ is the edge of the conduction (valence) band for n-type (p-type) SiNWs.  
Furthermore, 
for n-doped [110] SiNWs of diameter $2 R_{NW} = 1$ nm, they found that a donated electron de-localizes significantly along the wire axis, consistent with our results. 
But for thicker wires, they found 
$R_{loc}$ increasing from $R_{loc} = 2 \: \rm{\AA}$ for $2 R_{NW} = 1.5$ nm to $R_{loc} = 2$ nm in bulk.  For slightly thicker wires than the ones we model, localization could thus pose a challenge.  

For our transport calculations, we computed conductance by the Landauer formula as a (non-self-consistent) post-processing step \cite{Datta, Ke} to the Hamiltonian we calculated using periodic boundary conditions.  We used our in-house transport code, TIMES \cite{Georgios} to solve for the transmission function $T(E)$ from Green's functions for the Hamiltonians.  
This non-self-consistent approach is valid as a linear response to $V_{ds}$, but captures some non-Ohmic behavior because we integrate $T(E)$ rather than assume \cite{Datta} that $dT/dE \ll 1/{e V_{ds}}$.

\begin{figure}
\includegraphics[scale=0.28]{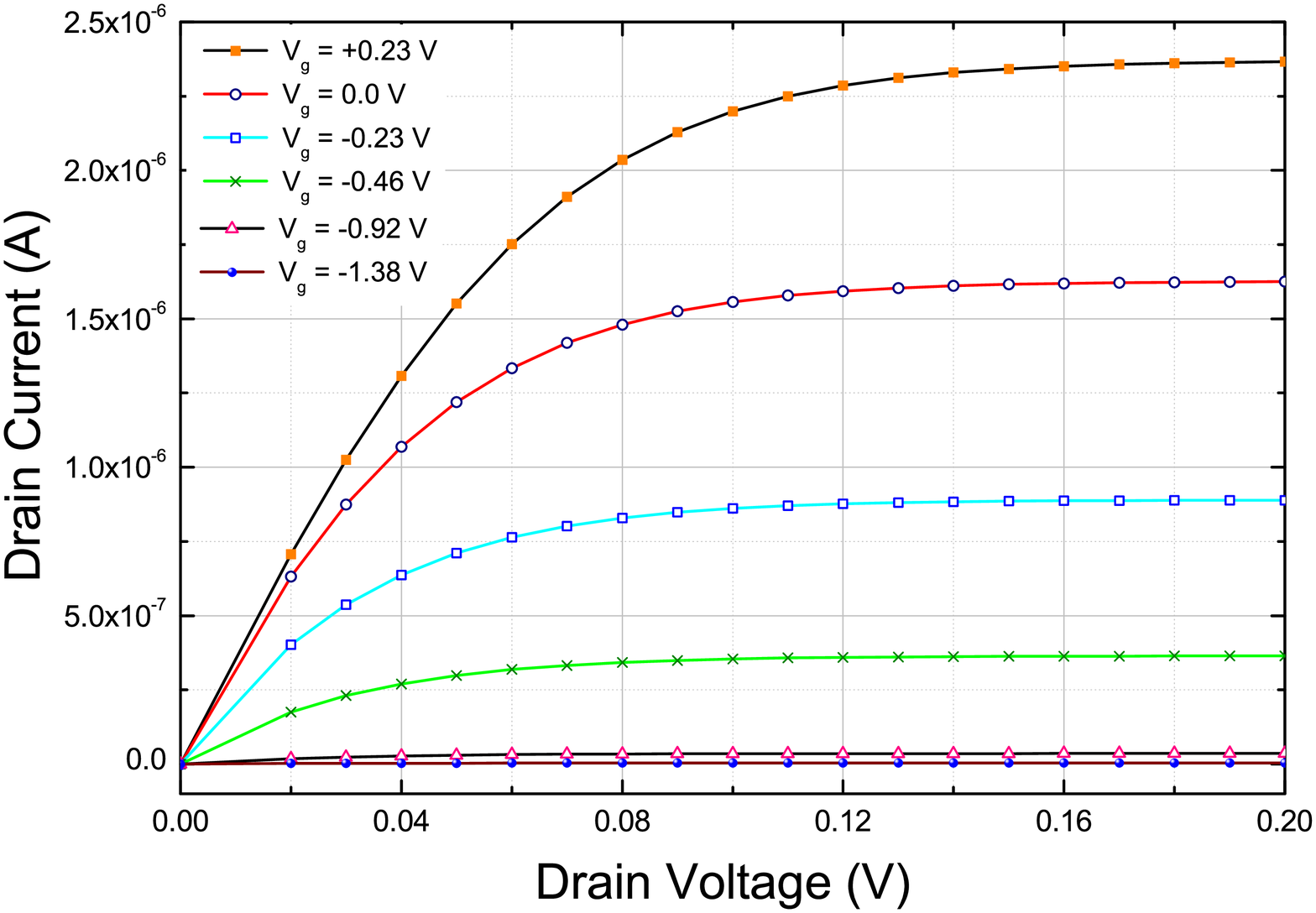}
\caption{\label{fig:IV-n}  $I$-$V_{ds}$ characteristic for SiNW junctionless transistor doped n-type by As atoms with dimensions shown in Fig.~\ref{fig:device}.  } 
\end{figure}
\begin{figure}
\includegraphics[scale=0.29]{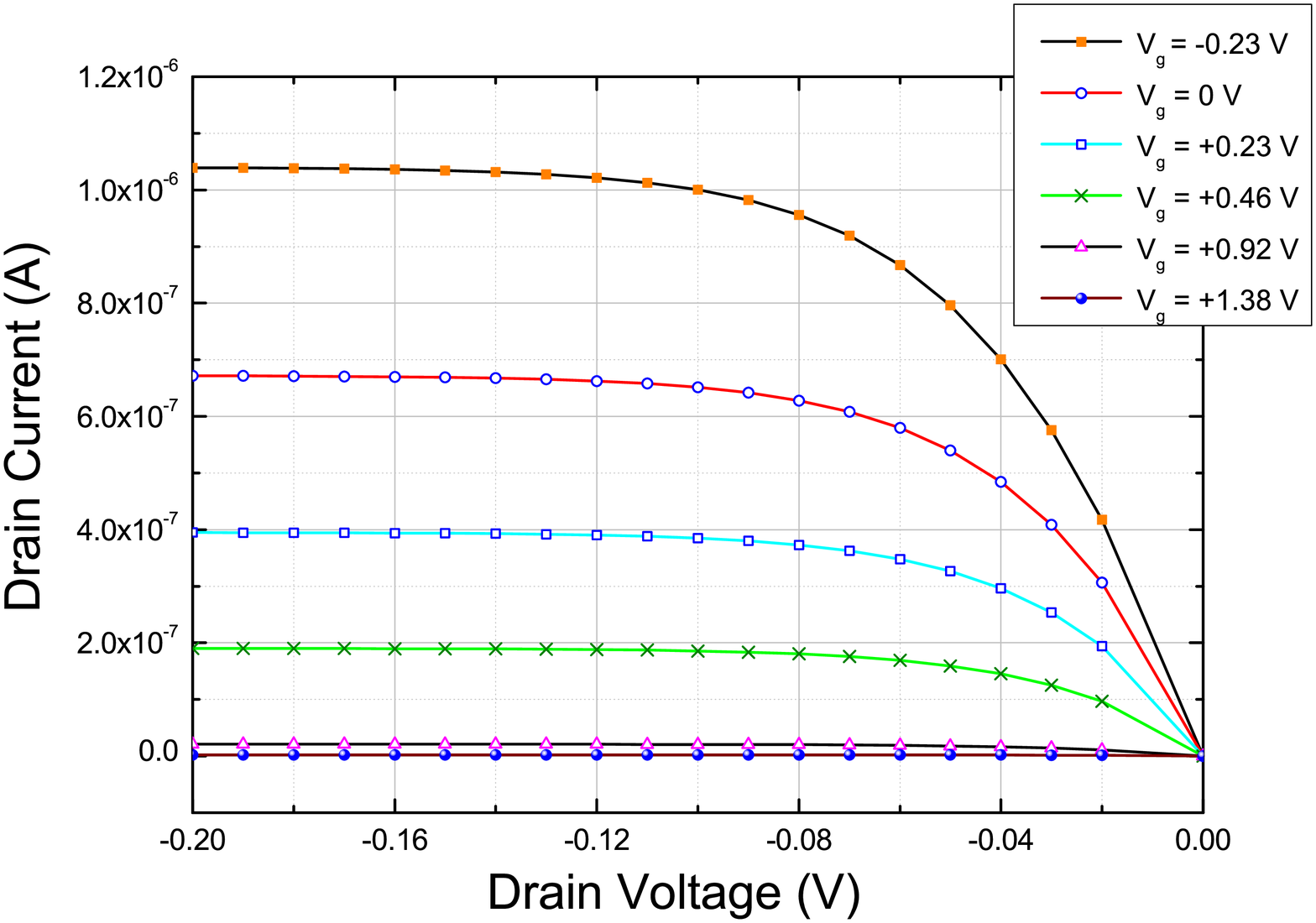}
\caption{\label{fig:IV-p} $I$-$V_{ds}$ characteristic for SiNW junctionless transistor doped p-type by Ga atoms.  }
\end{figure}

Figures \ref{fig:IV-n} and \ref{fig:IV-p} show the calculated $I$-$V_{ds}$ characteristics for n- and p-type junctionless devices, respectively.  Clearly, these devices are on for $V_g=0$ V, and they turn off based on a pinch-off principle when $V_g$ causes a sufficiently large barrier in the gating region.  Short-channel effects 
are a serious issue at these length scales, as tunneling across the $V_g$ barrier could undermine the device's effectiveness.  To mitigate short-channel effects, a rule of thumb for GAA geometry requires gate length \cite{ChanLength, JP-book} 
\[
L_g > 2 R_{NW},
\]  
a condition satisfied here by only a small margin. 

However, GAA geometries are well-known to have superior gate control, 
and were predicted to have nearly ideal subthreshold slopes for longer devices \cite{Cynthia}.  
Our results confirm even for gate lengths $\sim$3 nm, the junctionless GAA design has very good electrostatic control of the gate, enabling the devices to turn off.  
This is our most important finding \footnote{We validated this by using $V_g = 0$ V, but introducing a uniform energy shift to the NEGF Hamiltonian in the gated region.  This 
neglects capacitance and screening in the oxide and SiNW, but enables us to quantify the tunneling current through the channel.}.  

Our quantitative prediction of the turnoff gate voltage $V_{off}$ is an upper bound because of the lack of self-consistency in our NEGF calculations  
and the limited supercell.
Still, our calculations indicate a subthreshold slope 
close to the ideal, and much better than for other nanoscale device designs \cite{ChanLength, JP-book}.  

Kim and Lundstrom \cite{KimLund} analytically modeled junctioned SiNW MOSFETs of diameter a few nm, and found a similar saturation effect in 
the $I$-$V_{ds}$ characteristics.  Like them, we find 
its basic cause is the raising of the $\Gamma$-point in one terminal above the Fermi level in the other terminal at large $V_{ds}$, causing current saturation.  This effect is 
present even at $V_g = 0$.  


We found various effects of dopant positioning of relevance to prospective device design.  
First, as already mentioned, the donated electrons de-localize over a distance $R_{loc} \sim L_g$, so for 
\be
N \: \left( \pi R_{NW}^2 \; L_g \right) \gtrsim 1,
\label{eq:num-dopants-gate}
\ee
where $N$ is the doping concentration, the junctionless design is the only practical one that relies on local doping.  We note further that $N$ typically must {\em increase} with decreasing $L_g$ in order to maintain sufficient on current \cite{JP-book}.  

In addition, we found that the positioning of the dopant within the wire cross section makes a significant difference in the band structure of the device.  A periodic array of dopants near the SiNW surface is found to create a dopant band or otherwise narrows the SiNW band gap, which is ordinarily about twice the band gap of bulk Si \cite{Georgios}.  This narrowing leads to a steeper $I$-$V_{ds}$ characteristic.  



We have performed first-principles transport simulations on junctionless gate-all-around SiNW devices of radius $R_{NW}=0.6$ nm and gate length $L_g=3.1$ nm.  We predict that the junctionless transistor continues to work well at this scale, turning off with source-drain leakage $I_{off} < 10^{-6} \: I_{on}$, 
and has good electrostatic control and a good subthreshold characteristic.  By contrast, other designs with junctions or a single gate are unlikely to work at this scale.    Finally, dopant fluctuations may affect the band structure and various performance factors of the device.  But the basic operating principle of the junctionless SiNW is robust against dopant fluctuations.


We would like to thank J.~Andreas Larsson for useful discussions.  This research was funded by Science Foundation Ireland under grant 06/IN.1/I857.

\end{document}